\documentclass[lettersize,journal]{IEEEtran}
\usepackage{amsmath,amsfonts}
\usepackage{algorithmic}
\usepackage{algorithm}
\usepackage{array}
\usepackage{textcomp}
\usepackage{stfloats}
\usepackage{url}
\usepackage{CJKutf8}
\usepackage{verbatim}
\usepackage{graphicx}
\usepackage{cleveref}
\usepackage{cite}
\usepackage{utfsym}
\usepackage{subcaption}
\usepackage{booktabs}
\usepackage{mhchem}
\usepackage{amsmath,bm,flushend}
\usepackage{balance}

\usepackage{colortbl}
\usepackage{xcolor}
\usepackage{tikz}
\usepackage{etoolbox}
\newrobustcmd*{\mytriangle}[1]{\tikz{\filldraw[draw=#1,fill=#1] (0,0) --
(0.2cm,0) -- (0.1cm,0.2cm);}}

\begin{document}
\begin{CJK}{UTF8}{gbsn}
\title{DiffSG: A Generative Solver for Network Optimization with Diffusion Model} 

\author{Ruihuai~Liang, Bo~Yang,  Zhiwen Yu, Bin Guo, Xuelin~Cao,   
 M\'erouane Debbah, H. Vincent Poor,
 and Chau Yuen
\thanks{R. Liang, B. Yang, Z. Yu, and B. Guo are with the School of Computer Science, Northwestern Polytechnical University, Xi'an, Shaanxi, 710129, China.

X. Cao is with the School of Cyber Engineering, Xidian University, Xi'an, Shaanxi, 710071, China. 

M. Debbah is with the Center for 6G Technology, Khalifa University of Science and Technology, P O Box 127788, Abu Dhabi, United Arab Emirates. 

H. V. Poor is with the Department of Electrical and Computer Engineering, Princeton University, Princeton, NJ 08544, USA.

C. Yuen is with the School of Electrical and Electronics Engineering, Nanyang Technological University, Singapore. 



 }
}

\markboth{Journal of \LaTeX\ Class Files,~Vol.~14, No.~8, February~2025}%
{Shell \MakeLowercase{\textit{et al.}}: A Sample Article Using IEEEtran.cls for IEEE Journals}


\maketitle

\begin{abstract}
Generative diffusion models, famous for their performance in image generation, are popular in various cross-domain applications. However, their use in the communication community has been mostly limited to auxiliary tasks like data modeling and feature extraction. These models hold greater promise for fundamental problems in network optimization compared to traditional machine learning methods. Discriminative deep learning often falls short due to its single-step input-output mapping and lack of global awareness of the solution space, especially given the complexity of network optimization's objective functions. In contrast, generative diffusion models can consider a broader range of solutions and exhibit stronger generalization by learning parameters that describe the distribution of the underlying solution space, with higher probabilities assigned to better solutions. 
We propose a new framework Diffusion Model-based Solution Generation (D\scalebox{0.75}{IFF}SG), which leverages the intrinsic distribution learning capabilities of generative diffusion models to learn high-quality solution distributions based on given inputs. The optimal solution within this distribution is highly probable, allowing it to be effectively reached through repeated sampling. We validate the performance of D\scalebox{0.75}{IFF}SG on several typical network optimization problems, including mixed-integer non-linear programming, convex optimization, and hierarchical non-convex optimization. Our results demonstrate that D\scalebox{0.75}{IFF}SG outperforms existing baseline methods not only on in-domain inputs but also on out-of-domain inputs. 
In summary, we demonstrate the potential of generative diffusion models in tackling complex network optimization problems and outline a promising path for their broader application in the communication community. 
Our code is available at https://github.com/qiyu3816/DiffSG.
\end{abstract}


\section{Introduction}

\IEEEPARstart{D}{iffusion} generative models, renowned for their exceptional efficacy among generative models, have found extensive applications beyond image, speech, and video generation, encompassing scientific tasks such as graph generation and 3D structure generation \cite{chen2024opportunities}. However, their utilization in communication networks largely remains confined to data modeling, such as wireless channel coding \cite{sengupta2023generative} and feature extraction \cite{du2024enhancing}. These works typically apply Generative Diffusion Models (GDMs) either for data modeling and sampling or as feature extractors for other deep learning methods, such as reinforcement learning. There remains valuable potential for GDMs to be explored in communication networks, particularly for direct applications in solving network optimization problems.

Network optimization problems arise in many wireless networking applications, such as joint sensing, communication, computing and control \cite{shi2023machine}. This process generally involves developing optimal allocation plans using limited resources, to maximize or minimize an objective function under the given network parameters and specific constraints, which has the general format of $\min_{\mathbf{y}\in\mathcal{Y}} f(\mathbf{x},\mathbf{y}),\ s.t.\ g(\mathbf{x},\mathbf{y}) \leq 0,\ \rm{given}\ \mathbf{x}\in\mathcal{X}$. Network optimization problems generally exhibit three characteristics: complex objective functions, stringent constraints, and high-dimensional solution spaces. The objective functions may be non-differentiable or non-convex, and the constraints may include equality constraints, inequality constraints, and conditional constraints. High-dimensional inputs and outputs result in a huge solution space. Existing methods, whether traditional numerical approaches or machine learning techniques, typically aim to directly derive optimal solutions from inputs. However, traditional numerical methods demand intricate handcrafted designs, often possess high complexity, and lack transferability. Reinforcement learning requires manually designed reward and loss functions, where minor design deviations can significantly impact model convergence quality. Standard deep learning methods rely on pre-constructed ground truth datasets to learn mappings from inputs to solutions, but the complexity of the typical objective functions makes achieving a robust mapping quite difficult. Therefore, seeking an optimal solution in a large, non-differentiable, and constrained solution space in a single step is inherently infeasible, as this one-step approach lacks awareness of the current solution space. 

In contrast, learning to describe the solution space based on the input offers stronger global awareness, leading to better solutions. Specifically, the distribution of high-quality solutions within the solution space is such that the performance difference between high-quality solutions and the optimal solution on the objective function is negligible. In this distribution, the optimal solution has the highest probability, followed by high-quality solutions, while the probability of other non-high-quality solutions is almost zero. This concept of transforming a single output solution into a distribution of high-quality solutions within the solution space has been demonstrated in recent research \cite{qiu2022dimes,sun2023difusco,li2024distribution} and has yielded significant results in two classic combinatorial optimization problems: the Traveling Salesman Problem (TSP) and the Maximum Independent Set (MIS) problem. Notably, \cite{qiu2022dimes} first defined the parametrization of the solution space distribution, providing a continuously differentiable output target for neural network learning. 

From this general idea, we propose a new framework, Diffusion Model-based Solution Generation (D\scalebox{0.75}{IFF}SG), utilizing a GDM with intrinsic distribution learning capabilities \cite{li2024distribution} as the backbone. Diffusion models learn the inverse process (denoising process) of denoising random noise into the target distribution by reversing the forward process (noising process) that gradually adds noise to the data. This enables them to learn an unknown underlying distribution. Conditional generation can be easily implemented through \cite{ho2022classifier}, a variant of the classic Denoising Diffusion Probabilistic Models (DDPMs) \cite{ho2020denoising}. In D\scalebox{0.75}{IFF}SG, the noising process of the diffusion model acts as data augmentation for the optimal solution, facilitating the learning of the target high-quality solution distribution conditioned on the input. After training, better solutions can be obtained by parallel sampling from the high-quality solution distribution, as these better solutions have higher probabilities and thus a higher expectation of being sampled. When the high-quality solution distribution is correctly learned, a finite number of parallel samplings can reliably yield the optimal solution and ensure good performance on unseen inputs. 

There are a few works that have directly utilized generative models as optimization solvers. For instance, Large Language Models (LLMs) have been explored to solve differentiable simple constrained optimization and linear optimization problems, where the feasible solutions are iteratively generated and the optimal solution is approximated based on human feedback on the quality of the solutions at each step~\cite{yang2023large}. However, LLMs currently struggle with handling high-dimensional problems, and their performance is often disproportionate to the practical costs of training and inference. On the other hand, GDMs have also been investigated as optimization solvers. For instance, DIFUSCO \cite{sun2023difusco} and T2T \cite{li2024distribution} both employ graph GDMs to address TSP and MIS problems. However, these problems have relatively simple objective functions and constraints, and the application of such models to complex network optimization problems remains to be explored. Additionally, the authors of \cite{du2024enhancing} utilize the diffusion model to generate solutions for a purely convex optimization problem. Most of the aforementioned works do not address optimization problems from the perspective of learning high-quality solution distributions, and none fully realize the potential of GDMs as independent solution generators for network optimization problems. 

\textit{Different from the above works, we verify the feasibility and effectiveness of D\scalebox{0.75}{IFF}SG on several typical complex network optimization problems, laying a foundation for the application of GDMs in directly addressing network optimization problems.} Furthermore, by learning the underlying high-quality solution distribution directly from the data, D\scalebox{0.75}{IFF}SG is not constrained by the form of the objective function and constraints. This indicates that it is insensitive to properties such as differentiability, convexity, or whether the problem is discrete or continuous. Overall, this work presents a unique generative framework and opens up a new avenue for efficiently solving network optimization problems. 

\section{D\scalebox{0.75}{IFF}SG: A Generative Solver for Network Optimization}
In this section, we first explain the differences and advantages of learning a high-quality solution distribution for generation compared to learning an input-output mapping for single-step reasoning in neural network methods. Then we outline some basics of D\scalebox{0.75}{IFF}SG. Finally, a concrete implementation of D\scalebox{0.75}{IFF}SG is illustrated.

\subsection{Solution Distribution Learning or Input-output Mapping?}
\begin{figure*}[t]
\centering
\captionsetup{font={small}}
\includegraphics[width=12.5 cm]{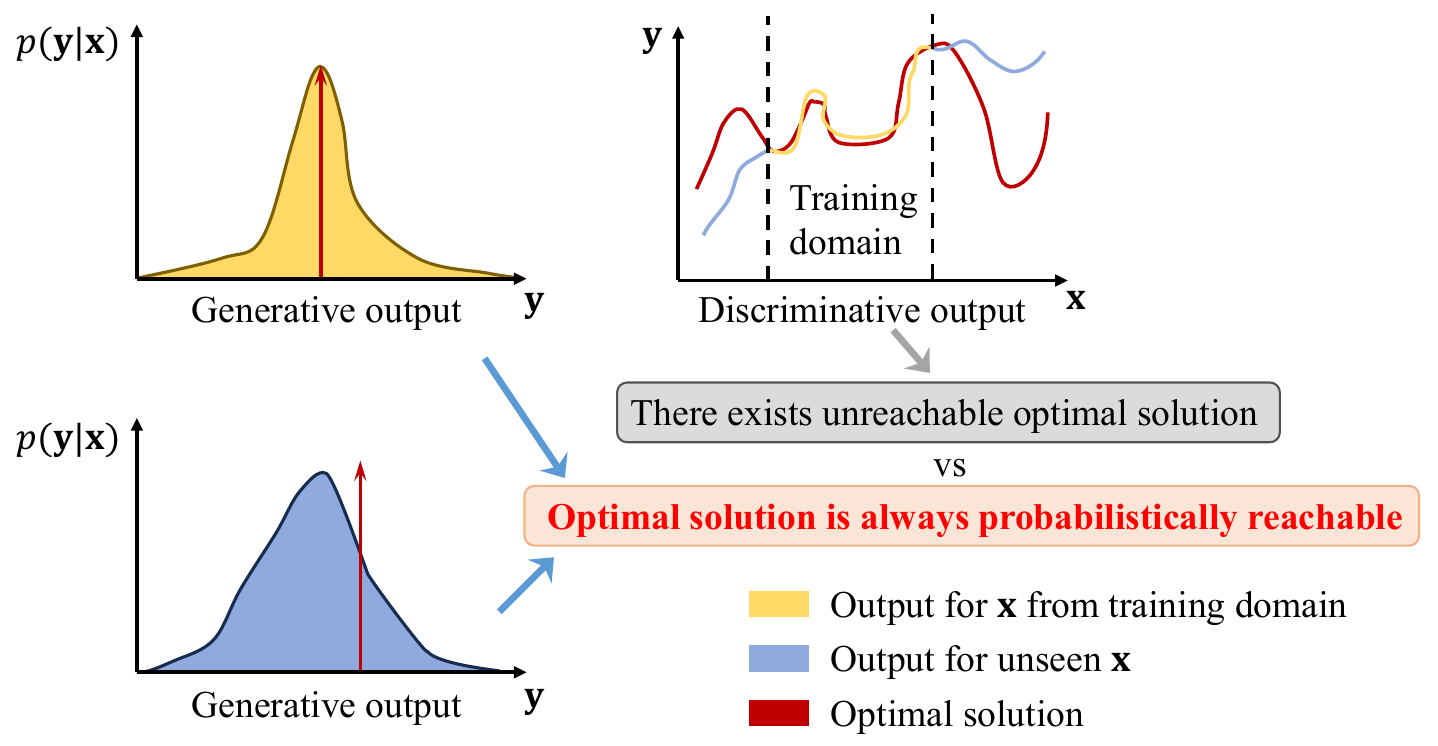}
\caption{Advantages of generative output over discriminative output.}
\label{fig_gen_dis_comp}
\end{figure*}

Generative models and discriminative models are two primary types of machine learning models. Generative models learn the joint distribution of samples $p(\mathbf{x},\mathbf{y})$ to generate novel data, while discriminative models learn the conditional distribution of samples $p(\mathbf{y}|\mathbf{x})$ to distinguish data. 

For the complex input-output relationships in network optimization, these two paradigms exhibit very different characteristics. As shown in Fig. \ref{fig_gen_dis_comp}, a discriminative model learns a fixed mapping from input to output. The complexity of network optimization problems (such as nonlinearity and non-convexity) can distort this end-to-end mapping relationship, as illustrated by the red curve in Fig. \ref{fig_gen_dis_comp}. {Even if the discriminative model has a small generalization error within the training domain, the generalization error outside the training domain will inevitably increase and become unpredictable.} This means that for a discriminative model, there will always be some inputs $\mathbf{x}$, especially those outside the training domain, for which the model output cannot reach or even approach the optimal solution. 

In contrast, a generative model learns the distribution $p(\mathbf{x},\mathbf{y})$ and outputs a sampled solution $\mathbf{y}$ that balances fidelity and diversity under the guidance of the implicit condition $p(\mathbf{y}|\mathbf{x})$. Here, fidelity refers to the quality and realism of the sampled solution conditioned on $\mathbf{x}$, and diversity refers to the variability of distinct sampled solutions that can be obtained by repeated sampling given $\mathbf{x}$. As shown in Fig. \ref{fig_gen_dis_comp}, high fidelity distinguishes generative solvers from purely random solvers. {With the synergy of diversity, the optimal solution for inputs within the training domain is highly likely to be reachable, and the optimal solution for inputs outside the training domain is also reachable with a certain probability gap. }

With this advantage, generative models can exhibit stronger generalization abilities on complex optimization problems compared to discriminative models, requiring only negligible parallel sampling. Consequently, we can transform the original problem of solving for the optimal output of the given input into an equivalent problem of learning a high-quality solution distribution, which can then be implemented using a generative model. 

\subsection{Basics of D\scalebox{0.75}{IFF}SG}
Inspired by \cite{qiu2022dimes}, we define the high-quality solution distribution as $p_{\boldsymbol{\theta}}(\mathbf{y}|\mathbf{x})\propto {\rm exp}(\sum^N_{i=1} y_i \theta_i),\mathbf{y}\in\mathcal{Y}_{\mathbf{x}},\ \mathbf{x}\in\mathcal{X}$, where $\boldsymbol{\theta}\in\mathbb{R}^N$ and $p_{\boldsymbol{\theta}}$ is an energy function indicating the probability of each solution over the feasible solution space $\mathcal{Y}_{\mathbf{x}}$. We suppose that the higher-quality solution in the distribution described by $\boldsymbol{\theta}$ should be with higher probability. Learning the parametrization of this high-quality solution distribution naturally enhances the model's global awareness of the solution space. There exists $\boldsymbol{\theta}^*$ that is the optimal parametrization of the solution space for a given $\mathbf{x}\in\mathcal{X}$ if $\boldsymbol{\theta}^*$ satisfies $p_{\boldsymbol{\theta}^*}(\mathbf{y}^*|\mathbf{x})=1$, where $\mathbf{y}^*$ is the optimal solution given $\mathbf{x}$. Obviously, even without achieving optimal parametrization, as long as the optimal solution has a high probability within the distribution, it is entirely feasible to reach the optimal solution through sampling. Therefore, our goal is to use a model to implicitly learn the $\boldsymbol{\theta}$ given $\mathbf{x}$ and sample from the corresponding solution distribution. Since the fundamental capability of generative models is to learn distributions and perform sampling, we can directly leverage GDM, one of the most effective and time-efficient generative models.

Diffusion models are a type of generative model that gradually adds noise to real data and learns to denoise it at each step. The result of the noising process is to convert the data into completely noisy data, such as by continuously adding standard Gaussian noise until the data becomes a pure Gaussian distribution. The model learns to denoise data at various noise levels, continuously refining the corrupted data until clean data is obtained. The essence of GDMs is to learn a data distribution $p(\mathbf{y})$ and achieve sampling from it. When a label $\mathbf{x}$ (sometimes referred to as a description or prompt) is available, the target data distribution is represented as the joint distribution $p(\mathbf{x},\mathbf{y})$, and it is possible to sample $\mathbf{y}$ from a target distribution conditioned on $\mathbf{x}$ through a conditional generation mechanism \cite{ho2022classifier}. In the context of our network optimization problem, solution is the data $\mathbf{y}$ being processed in the noising and denoising stages of the diffusion model, as illustrated in Fig. \ref{fig_DIFFSG_framework}.

Considering the complexity of the input-output mapping that a discriminative model needs to learn, it is foreseen that the high-quality solution distribution that the diffusion model needs to learn is also intractable. However, diffusion models are insensitive to the form of the target distribution, even when it is multi-modal, and therefore have the ability to overcome this challenge. A multi-modal distribution has a probability density function with multiple distinct peaks (e.g., a mixture of Gaussian and Laplace distributions) and its exact probability density function is often impractical to derive. Existing research indicates that in both theoretical evaluation and engineering verification, the generalization error of diffusion models is polynomially small rather than exponentially large with respect to the number of training samples and model capacity \cite{li2024generalization}. This avoids the curse of dimensionality and supports the application of diffusion models to a wider range of problems. GDMs have now also been proven to exhibit excellent properties in inference time scaling \cite{ma2025inference}, making them particularly well-suited for computationally intensive optimization problems in the networking domain.

Since the high-quality solution distributions corresponding to different $\mathbf{x}$ are distinct, $\mathbf{x}$ can be used as conditional guidance for generation. Specifically, the model implicitly learns the guidance of $P(\mathbf{y}|\mathbf{x})$ during training, effectively embedding a discriminative model within the generative neural network~\cite{ho2022classifier}. Additionally, the well-known high training and inference costs of generative models can be mitigated through accelerated sampling techniques. For example, Denoising Diffusion Implicit Models (DDIM) \cite{song2021denoising} significantly reduces the number of steps required for denoising and the number of inferences required by the model, by rewriting the sampling equation while maintaining denoising quality. 

Therefore, diffusion models provide a reliable method for learning complex high-quality solution distributions in network optimization. 

\subsection{Implementation}

\begin{figure*}[t]
\centering
\captionsetup{font={small}}
\includegraphics[width=15 cm]{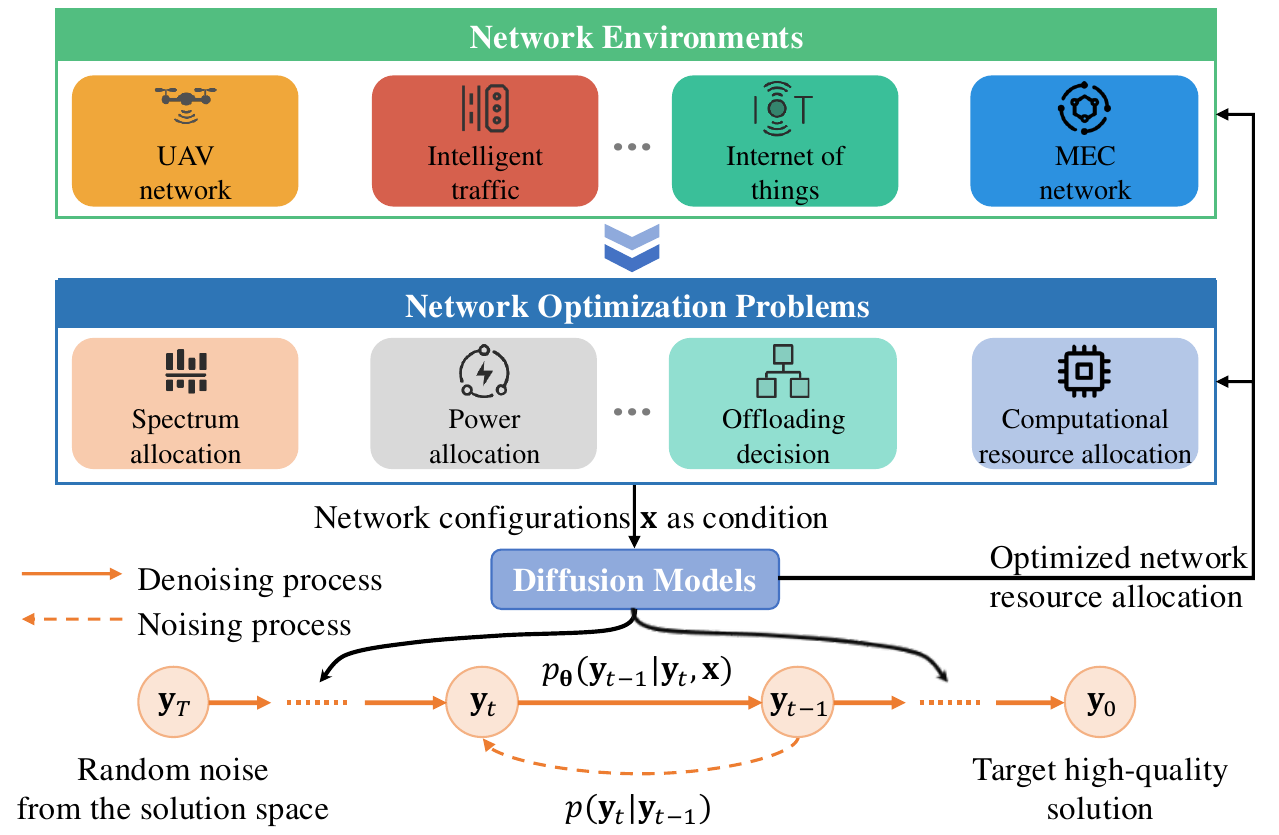}
\caption{The proposed D\scalebox{0.75}{IFF}SG framework.}
\label{fig_DIFFSG_framework}
\end{figure*}

As shown in Fig. \ref{fig_DIFFSG_framework}, the D\scalebox{0.75}{IFF}SG framework addresses network optimization problems arising from various network environments. The diffusion model is trained and used for sampling conditioned on network state information $\mathbf{x}$, such as channel gains and computational task parameters. During training, the model uses paired ground-truth examples $(\mathbf{x},\mathbf{y})$ for supervised learning. Based on a predefined number of diffusion steps $T$, the noise level is chosen uniformly and Gaussian noise is added to the true solution to obtain corrupted solutions with varying levels of noise. The model predicts the noise at each step for these corrupted solutions as $p_{\mathbf{\theta}}(\mathbf{y}_{t-1}|\mathbf{y}_t, \mathbf{x})$. The loss is calculated using the predicted noise and the real noise, and the model is updated to align its predictions closer to the actual noise.

During inference, samples are drawn from a known random distribution (random solutions in the solution space) as noisy solutions. The model then predicts and denoises these solutions step by step, starting from $\mathbf{y}_T$. Random noise is introduced into the denoising formula (sampling formula) to maintain sample diversity and smooth the distribution \cite{li2024generalization}. Throughout both training and inference, the model takes the condition $\mathbf{x}$ and the current solution $\mathbf{y}_t$ as inputs simultaneously. Intuitively, applying the diffusion model to network optimization involves modeling the entire feasible solution space and sampling solutions from the high-quality solution distribution within that space, where generated solutions are used to optimize the allocation of corresponding network resources.

Because the model only needs to learn the distribution from samples, these samples can originate from any form of optimization problem, making the framework insensitive to various characteristics of network optimization problems, such as differentiable and non-differentiable, convex and non-convex, linear and non-linear, discrete and continuous. This significantly enhances the generality of the proposed D\scalebox{0.75}{IFF}SG framework. The deployment paradigm we are currently developing is centralized. The model is typically deployed on a server node, executing real-time network optimization outputs based on current network state input parameters. In Fig. \ref{fig_DIFFSG_framework}, we illustrate common scenarios in real network environments. The model obtains network state parameters from specific scenarios and then inputs them as conditions to generate high-quality solutions. These solutions determine the allocation and scheduling of various physical resources, thereby achieving network optimization for specified objectives. 

\section{Case Study and Performance Evaluation}\label{sec_datasets}

\subsection{Overview of Cases and Models}
\begin{figure*}[t]
\centering
\captionsetup{font={small}}
\includegraphics[width=18cm]{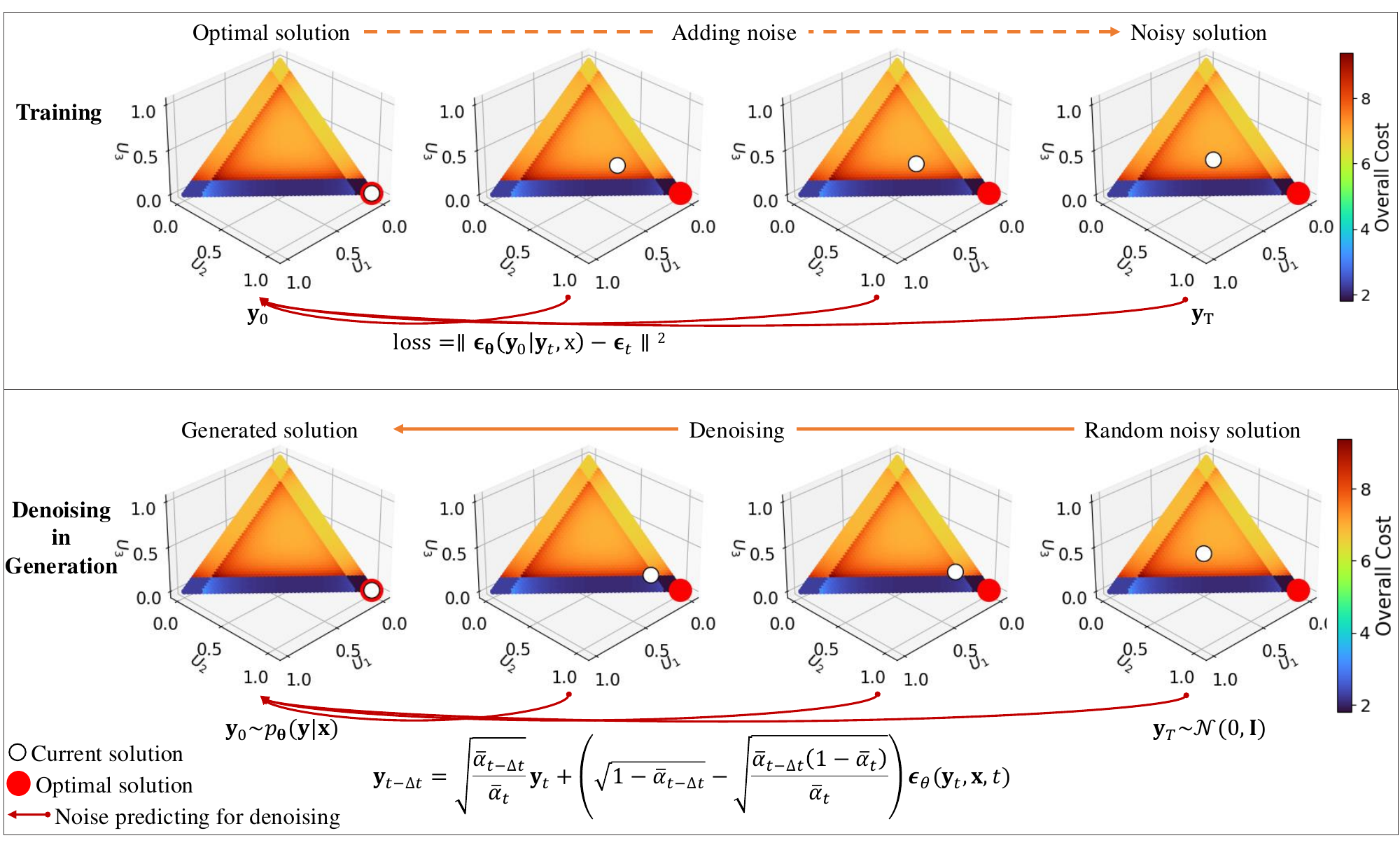}
\caption{The training and generation process for the computation offloading problem (CO), including the current solution and the optimal solution within the solution space determined by a given $\mathbf{x}$.}
\label{fig_co_diffusion_process}
\end{figure*}

\subsubsection{Case Problems}
We consider three typical network optimization problems for our case study: computation offloading (CO)~\cite{yang2020computation}, maximizing the sum rate of multiple channels (MSR)~\cite{du2024enhancing}, and maximizing the sum rate of multiple channels in the NOMA-UAV system (NU)~\cite{zhang2020joint}. We maintain consistency with the original work for modeling these three network optimization problems. 

For the CO problem, the input parameters include network and computing task parameters, while the output is the joint optimization of the offloading decision and computational resource allocation. The objective is to minimize the total weighted cost of latency and power consumption of the computing tasks. This belongs to an NP-hard mixed integer non-linear programming (MINLP) problem. Specifically, each optimization round of the CO problem is a binary offloading problem. The offloading decision variable is represented by a binary indicator (0 or 1), where 0 signifies local execution and 1 indicates offloading to the edge server. Meanwhile, computational resource allocation is represented by a continuous variable between 0 and 1, denoting the proportion of edge server resources allocated to the offloaded tasks. The total cost to be minimized consists of local execution cost, offloading transmission cost, and offloading execution cost. The given input parameters include task data size, channel gain, Orthogonal Multi-Access (OMA) \cite{yang2020computation} channel bandwidth, and the available computational resources at both the local and edge servers. To facilitate experimental validation, we consider a single-server, three-user setting, which avoids trivial solutions that all users execute tasks locally. The implementation directly derives offloading decisions from the resource allocation results (e.g., if the allocated computational resource is less than 0.1, the offloading decision is set to 0). Under this setup, the solution space is effectively a plane within a 3-D computational resource allocation space, where offloaded tasks fully utilize the available server resources, and the primary decision to be made is the allocation ratio. The D\scalebox{0.75}{IFF}SG model is deployed on the edge server to predict offloading decisions and resource allocation.

For the MSR problem, the inputs are the available total power and the gain parameters of each channel, and the output is the optimization of channel power allocation. The objective is to maximize the sum rate of channels, which is a common convex optimization problem. Specifically, the MSR problem considers a single-server, multi-channel setting in OMA communication. In each optimization round, the input parameters include the channel gains and the total available power at the server. The objective is to maximize the total transmission rate by optimally allocating power across the channels. Similarly, we consider two configurations: one with 3 channels and another with 80 channels, both following a power allocation strategy that fully utilizes all available transmission power. Under this setup, the corresponding solution space forms a bounded plane in the given dimension, which appears as a triangular plane in the three-dimensional channel power allocation space. The D\scalebox{0.75}{IFF}SG model is deployed on the server to allocate power.

For the NU problem, the inputs include ground terminal coordinates and network status, with the output being the joint optimization of the UAV coordinates and channel power allocation. The objective is to maximize the sum rate of channels, which is a hierarchical non-convex optimization problem. Specifically, the NU problem involves a Non-Orthogonal Multi-Access (NOMA) communication system consisting of a UAV and multiple ground terminals, to maximize the total transmission rate from all terminals to the UAV \cite{zhang2020joint}. The input parameters include the 2D ground locations of the terminals, the UAV's altitude, and its available communication power. The optimization variables are the UAV’s target 2D coordinate and the power allocated to each ground terminal’s channel. Similarly, we consider a setup with a single UAV and three ground terminals. As a result, the optimization variables include the UAV’s 2D coordinate and a 3D power allocation, which can be represented separately in a 2D and a 3D solution space. The D\scalebox{0.75}{IFF}SG model is deployed on the UAV for coordinate scheduling and power allocation.

\subsubsection{Datasets}
Our dataset is built following the settings of the original studies \cite{yang2020computation,du2024enhancing,zhang2020joint}. The datasets provide paired samples in the form of $(\mathbf{x}, \mathbf{y})$, where $\mathbf{x}$ is the input and $\mathbf{y}$ is the corresponding optimal solution. We use exhaustive methods and existing solution algorithms (only for MSR is feasible) to obtain the optimal solution for the dataset. Specifically, the original dataset sizes for these three experiments are $50,000$ samples for CO, $10,000$ samples for MSR (3 channels), $10,000$ samples for MSR (80 channels), and $10,000$ samples for NU, respectively.

To demonstrate the superior robustness of the generative model over the discriminative model on Out-Of-Domain (OOD) input data, as described in Fig. \ref{fig_gen_dis_comp}, we construct OOD validation datasets. For CO, the task data size and local computational resources exceed the training domain; for MSR, the channel gain range and total available server power go beyond the training domain; and for NU, the UAV’s available communication power extends beyond the training domain. The extent to which these input variables exceed the training range varies from a minimum of 20\% to a maximum of 100\% (e.g., the average task data size in CO increases from $2.5\times 10^{5}$ Bytes to $3\times 10^{5}$ Bytes, and the total power for MSR with 3 channels increases from 10W to 20W). The validation set contains $2,000$ samples for CO, $2,000$ for MSR (3 channels), $2,000$ for MSR (80 channels), and $2,000$ for NU.

\subsubsection{Baselines and Model Settings}
As an exploratory case study, we compare the proposed D\scalebox{0.75}{IFF}SG with three baseline methods: (1) Gradient Descent (GD), which is a simple numerical method and the Lagrange multiplier method is used to integrate the objective function and constraints into the gradient formulation; (2) Multi-Task Feedforward Neural Network (MTFNN), which is a typical discriminative machine learning method \cite{yang2020computation}; and (3) Proximal Policy Optimization (PPO), which is a deep reinforcement learning method for learning unimodal action distribution. In the experiment, D\scalebox{0.75}{IFF}SG applies the classifier-free conditional guidance mechanism \cite{ho2022classifier} and adopts the classic DDPM \cite{ho2020denoising} model. The noise schedule $(\alpha_t)^T_{t=1}$ is constructed by the cosine schedule that has been widely adopted in diffusion models \cite{ho2020denoising}. Denoising Diffusion Implicit Models (DDIM) \cite{ho2022classifier} acceleration from $\mathbf{y}_t$ to $\mathbf{y}_{t-\Delta_t}$ denoising is also supported in sampling. Notably, the maximum noising step $T$ of our model is set to $20$, which is significantly fewer than the thousands of steps typically used in DDPM for image generation tasks. Additionally, the conditional strength parameter $\omega$ is set to $500$, this is much higher than the value used for image generation in \cite{ho2022classifier} (less than 4). We will analyze these two unusual settings in detail in the next section.

\subsubsection{Metrics}
We define the performance metric, $\rm{exceed\_ratio}=\frac{f(\mathbf{x},\mathbf{y}_{pred})}{f(\mathbf{x},\mathbf{y}^*)}$, as the ratio of the output solution to that of the ground truth, as shown in Table \ref{tab_performance}. We can observe that the closer the $\rm exceed\_ratio$ is to $1$, the better the performance is achieved. 

\begin{figure*}[ht]
\centering
\captionsetup{font={small}}
\includegraphics[width=18cm]{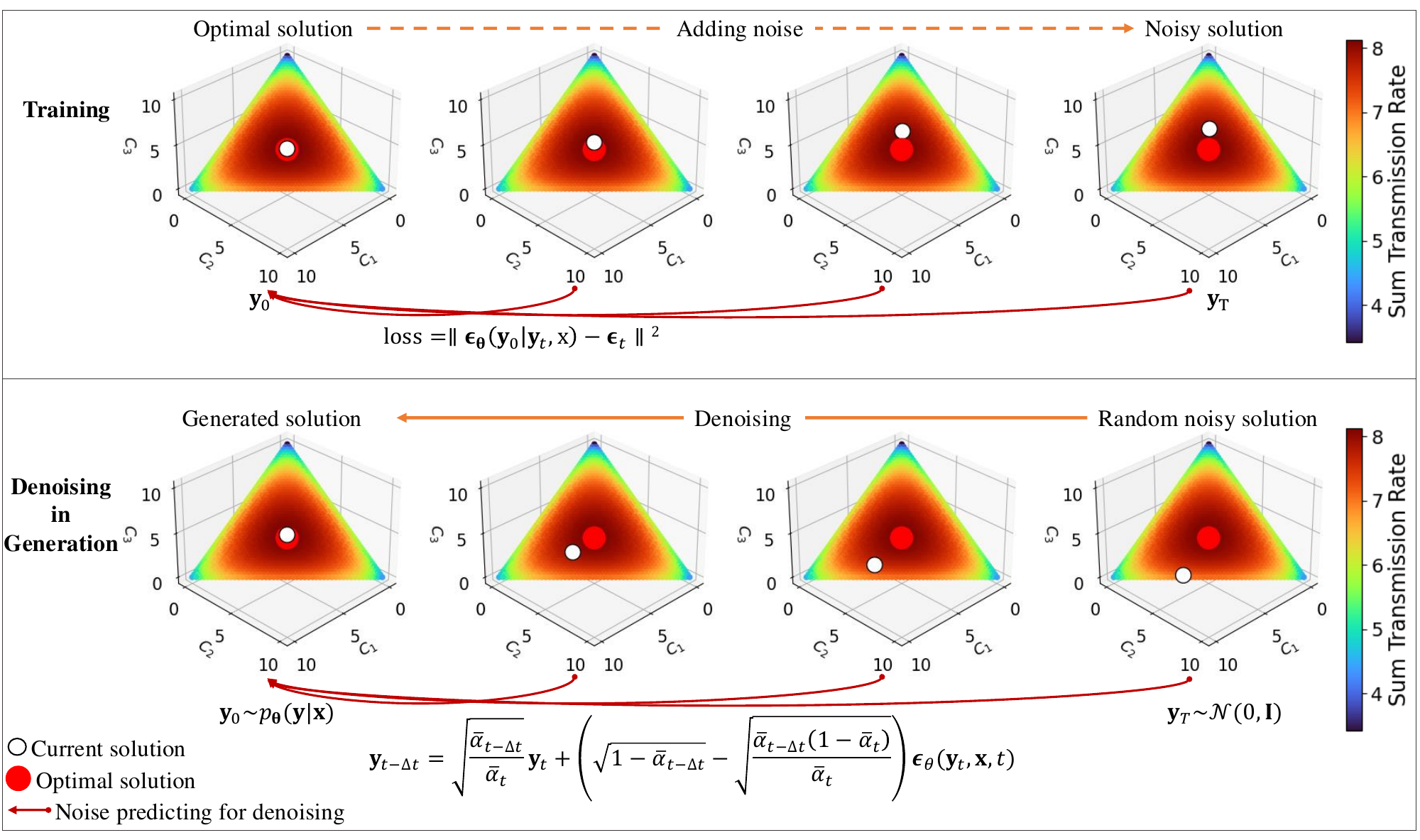}
\caption{The training and generation process for maximizing the sum rate of multiple channels (MSR), including the current solution and the optimal solution within the solution space determined by a given $\mathbf{x}$.}
\label{fig_msr_diffusion_process}
\end{figure*}

\subsection{Verification of Optimal Solution Generation }

\begin{figure*}[ht]
\centering
\captionsetup{font={small}}
\includegraphics[width=18cm]{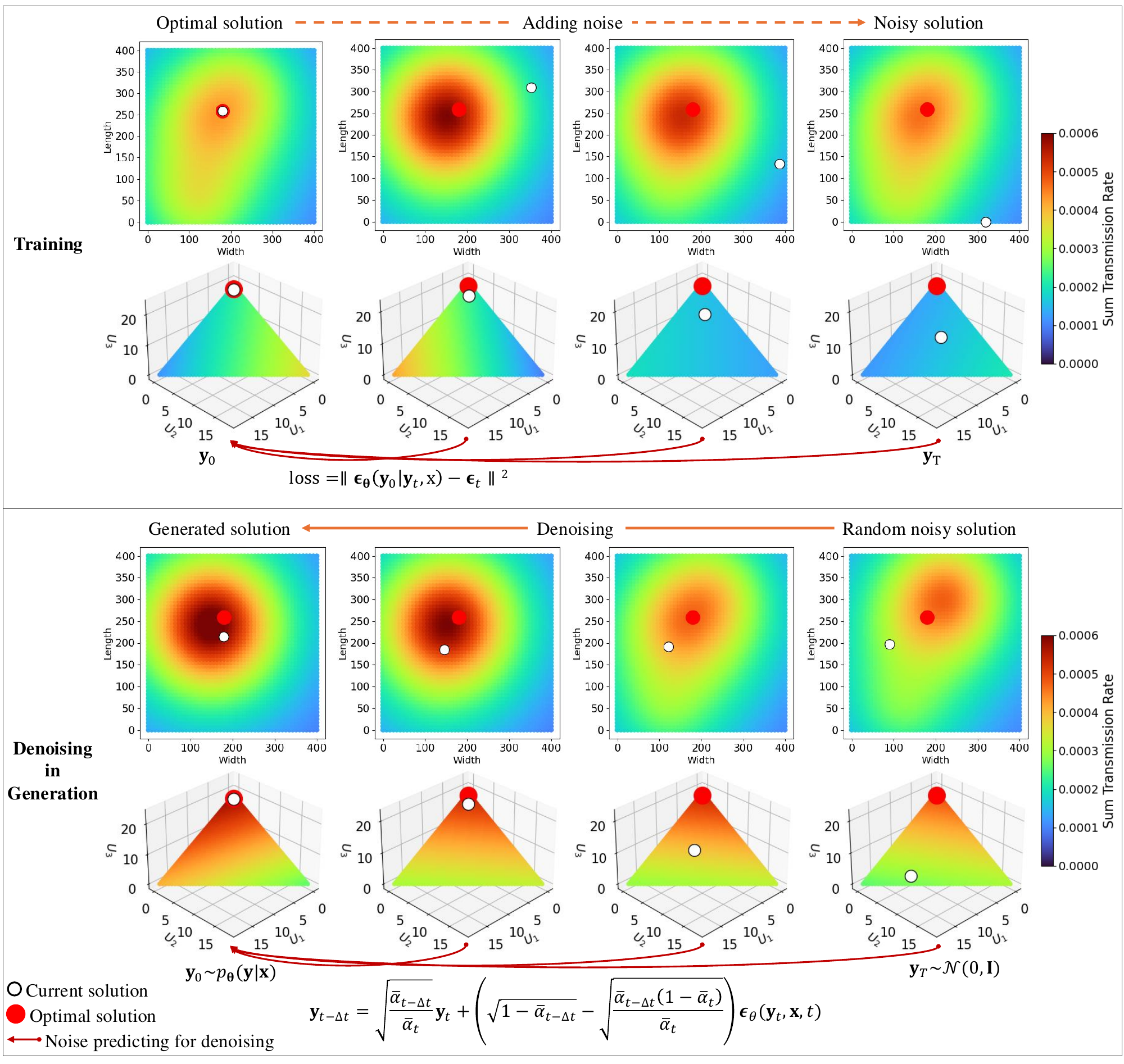}
\caption{The training and generation process for maximizing the sum rate of multiple channels in the NOMA-UAV system (NU), including the current solution and the optimal solution within the solution space determined by a given $\mathbf{x}$. The two optimization variables, the UAV's 2D coordinates and the channel power allocation for the three users, are displayed using a 2D heatmap and a 3D heatmap, respectively. The 2D heatmap at step $t$ is determined by the channel power allocation of $\mathbf{y}_t$, while the 3D heatmap is determined by the UAV's coordinates of $\mathbf{y}_t$.}
\label{fig_nu_diffusion_process}
\end{figure*}

The input variable $\mathbf{x}$ determines the form of the solution space for the objective function. We illustrate the solution spaces for each problem based on several different inputs $\mathbf{x}$. Each point in the solution space represents a feasible solution $\mathbf{y}$, and the color of the point represents the objective function value at that position. The white circles in the figure represent the solutions in the adding noise or denoising process, while the red circles indicate the optimal solution of given $\mathbf{x}$. In Figs. \ref{fig_co_diffusion_process}, \ref{fig_msr_diffusion_process}, \ref{fig_nu_diffusion_process}, we illustrate the training and generation processes for the CO, MSR (3 channels), and NU problems, each given a specific input $\mathbf{x}$. These figures demonstrate the dynamics of solutions during the noise adding and denoising process, as well as the convergence behavior of sampling. Specifically, during training, the model learns to predict the noise $\mathbf{\epsilon}_{\mathbf{\theta}}(\mathbf{y}_0|\mathbf{y}_t,\mathbf{x})$ required to recover the optimal solution from any given noisy solution $\mathbf{y}_t$ and calculates the loss based on the difference from the true noise $\mathbf{\epsilon}$. During sampling, the model directly denoises a completely random initial solution $\mathbf{y}_T$, which can be performed step by step or accelerated using DDIM sampling \cite{song2021denoising}.

Fig. \ref{fig_co_diffusion_process} shows the training and generation process of D\scalebox{0.75}{IFF}SG for the CO problem, where each dimension represents the proportion of computational resources allocated by the edge server to each user. The objective function here is to minimize the overall cost, so the bluest area with low values represents high-quality solutions. Considering joint optimization of offloading decision and computational resource allocation is a mixed-integer problem, truncated function values appear in the solution space. In Fig. \ref{fig_co_diffusion_process}, the noise adding process is illustrated using the same Gaussian noise for demonstration. However, during actual training, a new Gaussian noise is resampled at each step $t$, meaning that the noise direction of $\mathbf{y}_t$ relative to the optimal solution may vary across different $t$. The denoising process in Fig. \ref{fig_co_diffusion_process} successfully refines a random initial solution $\mathbf{y}_T$ step by step, ultimately converging to the optimal solution.

The problem of MSR is illustrated in Fig. \ref{fig_msr_diffusion_process}, using the same settings as Fig. \ref{fig_co_diffusion_process}, where each dimension represents the power allocated by the server to each channel. Here, the power is allocated to maximize the sum transmission rate, making the reddest area in the solution space represent high-quality solutions. Although the objective function of MSR, which has been studied extensively by \cite{du2024enhancing}, is smooth and convex, we conduct verification experiments on this problem and achieve effective convergence, where more in-depth principle and more robust generalization are studied.

The problem of NU is shown in Fig. \ref{fig_nu_diffusion_process}, which is significantly different from the first two problems. Here, the two axes of the 2D heatmap corresponding to the width and height of the UAV's geographical region plane, while each dimension of the 3D heatmap represents the power allocated by the UAV to each channel with ground terminals. In both 2D and 3D heatmap, the reddest area indicating high-quality solutions. In this hierarchical non-convex optimization, UAV coordinates and power allocation determine each other's solution space, making it meaningless to display the solution spaces of them collectively. Since any change in either the UAV's coordinate or its power allocation affects the solution space of the other variable, the heatmap patterns in Fig. \ref{fig_nu_diffusion_process} continuously evolve. In this case, the model implicitly learns conditional guidance from $\mathbf{x}$ to the target distribution of complete high-quality solutions. The simultaneous convergence of UAV coordinates and power allocation can be observed in Fig. \ref{fig_nu_diffusion_process}. 

In our experiments, the number of diffusion steps $T$ was consistently set to $20$. However, in the three figures above, we only select solutions at denoising steps $t=18,17,16,0$. It can be observed that D\scalebox{0.75}{IFF}SG converges faster than $T$, primarily due to the larger conditional strength parameter. Additionally, we set $T$ significantly lower than in typical image generation tasks, which can be discussed in the context of the recent inference time scaling concept \cite{ma2025inference}. Since the complexity of our network optimization problems is lower than that of high-resolution image generation, a high degree of inference time scaling—i.e., an excessive number of diffusion steps—is unnecessary for achieving convergence.

In the original classifier-free image generation \cite{ho2022classifier}, the condition strength factor $\omega$ was empirically set to less than $4$, and increasing it would lead to worse diversity. However, in network optimization problems, we have found that a small value of $\omega$ leads to insufficient conditional guidance strength, resulting in final outputs that are too random to converge accurately. The optimal value of $\omega$ in practice is much larger than the previous empirical value. We believe there are two possible reasons for this behavior. On one hand, the network optimization condition $\mathbf{x}$ is more difficult for the model to learn than text or categorical conditions for image generation. This is because changes in numerical features at the decimal level are much more challenging to capture and understand than changes in semantics or categories. This causes the conditional noise amplitude learned by the D\scalebox{0.75}{IFF}SG model from conditional training to be not strong enough. On the other hand, the goal of network optimization does not need to consider the diversity required in image generation. If the denoising conditions are set based on experience in other domains, the diversity of the final generated solutions will be excessively high, and the solution accuracy will decrease. Therefore, a large value of $\omega$ indicates low diversity for image generation but can bring gains for network optimization problems that require more determinism.

In summary, the results in Fig. \ref{fig_co_diffusion_process}, \ref{fig_msr_diffusion_process}, and \ref{fig_nu_diffusion_process} provide insights into the impact of relevant hyperparameters on the performance of solution generation for network optimization. More importantly, we demonstrate that D\scalebox{0.75}{IFF}SG effectively learns high-quality solution distributions and achieves convergence in solution generation across various problem types, rather than being limited to a specific class of problems. This demonstrates the reliable effectiveness and valuable prospects of GDMs for solution generation in dealing with network optimization problems.

\subsection{Performance and Generalization Evaluation}

\subsubsection{In-domain Evaluation}
In the experiments, MTFNN and PPO are both neural network methods, and we set their training dataset, epochs, learning rate, and optimizer to be consistent with D\scalebox{0.75}{IFF}SG. Due to the independence of each sample in the dataset, the PPO environment is simulated to have only one initial state and one final state during training, so the target solution is the action to be learned. It is observed from Table \ref{tab_performance} that D\scalebox{0.75}{IFF}SG outperforms both MTFNN and PPO on experimental problems in original domain, except for the simplest 3-channel MSR, where it achieves almost the same performance. Due to the supervised training feature, the MTFNN learns the one-step mapping relationship from the dataset, which yields limited results. In contrast, PPO assumes that the action distribution is an unimodal Gaussian, and the learning target cannot be predefined because the distribution form of the target solution is unknown. This highlights the advantage of diffusion models in generating any form of target distribution, which is beneficial in network optimization. 

Moreover, GD is the only non-neural network method considered. Conventionally, GD uses the gradient of the current solution on the objective function as the optimization direction, ultimately achieving the extreme value. For the three constrained problems here, we use the classical Lagrange multiplier method to convert all constraints into corresponding cost functions and merge them with the optimization objectives for gradient descent, with the final constrained solution achieved by adjusting the Lagrange multipliers. For the mixed-integer case of CO problem, the binary solution can control the objective function through split weighting (e.g., if $D$ is a binary output of 0-1, then $D$ can weigh the offloading cost, and $1-D$ can weigh the non-offloading cost), and the final solution can be clipped to obtain a valid solution. However, it is challenging for Lagrange multipliers to achieve perfectly accurate results, so GD's performance on CO and NU is limited except for the fully convex MSR cases. Moreover, NU is a hierarchical non-convex optimization problem. Even when NU is transformed into an unconstrained smooth objective function, vanilla gradient descent can easily fall into a local optimum. D\scalebox{0.75}{IFF}SG consistently achieves better performance than other neural baselines in these three network optimization problems of varying types and difficulties. 

\begin{table}[t]
\centering
\captionsetup{font={small}}
\caption{Performance comparison of different methods on case problems based on the $\rm{exceed\_ratio}$ metric, where values closer to 1 indicate better performance. The results include evaluations on both the original training domain and the OOD validation set (separated by '\textbar').}
\begin{tabular}{l@{\hskip 0.04in}c@{\hskip 0.08in}c@{\hskip 0.04in}c@{\hskip 0.04in}c}\label{tab_performance}
 & CO$\downarrow$ & MSR (3 channels)$\uparrow$ & MSR (80 channels)$\uparrow$ & NU$\uparrow$ \\
\midrule
GD & 1.30\textbar1.34 & 0.99\textbar0.99 & 0.92\textbar0.94 & 0.83\textbar0.85 \\
MTFNN & 1.67\textbar1.71 & 0.99\textbar0.99 & 0.89\textbar0.93 & 0.84\textbar0.82 \\
PPO & 1.60\textbar1.58 & 0.99\textbar0.99 & 0.85\textbar0.87 & 0.79\textbar0.53 \\
\rowcolor{green!40} D\scalebox{0.75}{IFF}SG & \textbf{1.03\textbar1.02} & \textbf{0.99\textbar0.99} & \textbf{0.93\textbar0.95} & \textbf{0.92\textbar0.90} \\
\bottomrule
\end{tabular}
\end{table}

\subsubsection{Out-of-domain Evaluation}
To validate and support the concept in Fig. \ref{fig_gen_dis_comp} that generative models exhibit greater robustness in solution optimization for OOD inputs compared to discriminative models, we evaluate the trained models using the OOD validation dataset described in Sec. \ref{sec_datasets}-A-2. As outlined in Sec. \ref{sec_datasets}-A-2, the input parameters $\mathbf{x}$ in these OOD datasets exceed the training domain to varying degrees. For different types of network optimization problems, this deviation can lead to unpredictable changes in the solution space and the location of the optimal solution.

As shown in Table \ref{tab_performance}, different methods exhibit varying levels of solution optimization quality on the OOD validation set for the same problem. These differences are primarily influenced by the diverse problem characteristics and underlying methodological principles. Therefore, the impact of each method on OOD input is relatively random. We mainly compare the magnitude of change and the final results.

Notably, despite all neural network-based methods employing the same data preprocessing and normalization techniques, D\scalebox{0.75}{IFF}SG consistently outperforms all baselines in generating high-quality solutions on the OOD validation set. Moreover, the quality of its solutions remains nearly identical to that on the in-domain validation set. This provides strong empirical evidence supporting the concept in Fig. \ref{fig_gen_dis_comp} that generative models offer superior generalization robustness to OOD inputs in network optimization problems. We believe that this is due to the unique high-quality solution distribution learning objective, and the stronger inference time scaling brought by the denoising process \cite{ma2025inference} than the discriminative model.

\subsubsection{Summary}
D\scalebox{0.75}{IFF}SG's robust generalization capability on OOD inputs offers significant advantages for applications in dynamic and time-varying network environments, reducing the impact of input drift on model performance and operational costs. We believe this will pave the way for broader applications of generative models and also the standardization of next-generation networked intelligence.

\subsection{Complexity Analysis}
D\scalebox{0.75}{IFF}SG currently adopts a classic U-Net neural network architecture \cite{ho2020denoising}, which we have restructured to transition from an image-output task to a vector-output task. As a result, the network no longer contains attention layers or convolutional layers. Specifically, the model embeds an input solution vector of $N$ dimensions into an $h$-dimensional vector, and then applies $n$ rounds of down-sampling and up-sampling using linear layers within the U-Net structure \cite{ho2020denoising}.

Regarding the parameter complexity of D\scalebox{0.75}{IFF}SG, the total number of model parameters is limited to $O(nh)$. Furthermore, given that the dimension of the conditional input vector $\mathbf{x}$ is $C$, and the time embedding vector dimension is empirically set to $4h$, the time complexity of D\scalebox{0.75}{IFF}SG is primarily determined by the residual blocks composed of linear layers, which dominate both the down-sampling and up-sampling operations. As a result, the total time complexity is constrained to $O(Nh+Ch+h^2)$. Among the terms, $h^2$ has a relatively large linear constant, while $Nh$ and $Ch$ have small linear constants. The down-sampling depth $n$ cannot be increased infinitely due to U-Net's down-sampling characteristic (upper-bounded by $h$), and since it appears in an exponent within the denominator, its contribution is negligible compared to constant terms in our setting. Consequently, the overall computational complexity exhibits a linear relationship with the number of diffusion steps $T$, i.e., $O((Nh+Ch+h^2)T)$.

In our experiments conducted on an Apple M2 chip with 24GB of memory, we set the model parameters as follows: for CO, MSR (3-channel), and MSR (80-channel), we used $h=64$ and $n = 4$, while for NU, we set $h = 32$ and $n = 3$. The average inference time per sample is 110.0ms, 99.1ms, 101.2ms, and 66.7ms, respectively. These results align with our theoretical analysis, where inference time is primarily influenced by the hidden vector dimension and model depth rather than the input or conditional vector dimensions.

Through both theoretical and engineering complexity analysis, we conclude that D\scalebox{0.75}{IFF}SG is highly feasible for network optimization applications. Its efficiency can be further improved by incorporating advanced acceleration techniques \cite{song2021denoising}. D\scalebox{0.75}{IFF}SG is comparable in complexity to other neural network methods, but the quality of its output solutions is superior. Moreover, compared to machine learning-assisted numerical algorithms or swarm-based methods, these approaches may achieve remarkable optimization performance on specific network tasks but are often constrained by labor-intensive expert design and excessive computational complexity. Although its complexity may not outperform classical algorithms in well-studied problems, D\scalebox{0.75}{IFF}SG demonstrates strong generalizability, making it particularly valuable for solving complex network optimization tasks where mature solutions are not yet available.

\section{Conclusion and Future Directions}
In this paper, we have introduced a novel network optimization framework, called D\scalebox{0.75}{IFF}SG, to generate high-quality solutions for network optimization problems. We have demonstrated that the proposed D\scalebox{0.75}{IFF}SG effectively converges across various optimization problems by utilizing a DDPM, transforming the goal of directly inferring the optimal solution into fitting a high-quality solution distribution. We have also seen that network optimization characteristics necessitate customized designs for diffusion models. Additionally, we have also successfully demonstrated the robust generalization of GDMs on OOD inputs, enabling their application in dynamic and time-varying network environments. These findings highlight the potential of GDMs in solution generation for network optimization, due to their high probability of reaching the optimal solution, the looseness of constraints on the form of the objective function, and their insensitivity to the distribution characteristics of the target solution. 

For our future work, several promising directions have the potential to be further explored. The current model we use is significantly smaller in scale compared to models in other well-established domains (less than a million parameters versus billions), indicating substantial potential for scalability. Additionally, while our experiments have primarily focused on validating effectiveness, further improvements require collecting state-of-the-art baselines across various approaches (e.g., numerical algorithms, reinforcement learning) or testing deployment in real-world environments. Regarding deployment strategies, our case problems currently assume a centralized deployment. However, given the diverse distributed scenarios in network environments, studying the distributed deployment of GDMs is also essential. Furthermore, considering the different data structures of various resources in network optimization, generative approaches such as discrete solution generation and structured data generation will play a crucial role in practical applications. At the same time, it is also attractive to consider improving the conditions for guided generation, explicitly integrating the objectives and constraints of network optimization to enhance versatility and interoperability.

Last but not least, while theoretical exploration and model mechanism analysis are inherently challenging, there remains significant research space for investigating the error and robustness of GDM-generated solutions through both theoretical and empirical studies.

\balance
\bibliographystyle{IEEEtran}

\end{CJK}
\end{document}